\documentclass[aps,prd,twocolumn,showpacs,showkeys,superscriptaddress]{revtex4-1}
\usepackage{latexsym}
\usepackage{amssymb}
\usepackage{amsmath}
\usepackage{amscd}
\usepackage{amsthm}
\usepackage{graphicx}
\usepackage{textcomp}
\usepackage{colortbl}
\usepackage{hyperref}
\usepackage[font={footnotesize,it}]{caption}
\usepackage{multirow}
\usepackage[T1]{fontenc}
\usepackage{ae,aecompl}

\begin{document}

\title{Constraining An Exact Brans-Dicke gravity theory with Recent Observations}
\author{Hassan Amirhashchi}
\email[]{h.amirhashchi@mhriau.ac.ir, hashchi@yahoo.com}
\affiliation{Department of Physics, Mahshahr Branch, Islamic Azad University,  Mahshahr, Iran}
\author{Anil Kumar Yadav}
\email[]{abanilyadav@yahoo.co.in}
\affiliation{Department of Physics, United College of Engineering and Research,Greater Noida - 201310, India}

\begin{abstract}
In this paper first we study Brans-Dicke equations with the cosmological constant to find an exact solution in the spatially flat Robertson-Walker metric. Then we use Observational Hubble data, the baryon acoustic oscillation distance ratio data as well as cosmic microwave background data from Planck to constrain parameters of the obtained Brans-Dicke model. To compare our results and find out the amount of deviation from general relativity, we also constrain concordance cosmological model using the same data.  In our theoretical model the Brans-Dicke coupling constant is replaced by a , say new, parameter namely \textgravedbl scalar field density $\Omega_{\phi}$\textacutedbl. Therefore, as $\Omega_{\phi}\to 0$ which is equivalent to $\omega\to \infty$ general relativity is recovered. In general, we found no significant deviation from general relativity. Our estimations show $\Omega_{\phi}=0.010^{+0.021}_{-0.012}$ which is equivalent to $\omega>1560$ at $95\%$ confidence level. We also obtained constraint the rate of change of gravitational constant, $\dot{G}/G$, at present time as $1.150\times 10^{-13}yr^{-1}<\dot{G}/G<1.198\times 10^{-13}yr^{-1}$ (at $1\sigma$ error). The total variation of gravitational constant, since the epoch of recombination, is also constrained as $-0.0084<\delta G/G<-0.0082$ at $68\%$ confidence level. 

\end{abstract}



\maketitle
\section{Introduction}
\label{sec:intro}
The concordance cosmological model ($\Lambda$CDM) of universe is credited today as the most simplest and successful cosmological model that describes the dynamics of present universe with acceleration but it suffers some fundamental problems on theoretical ground, which are reported by numerous cosmologists \cite{ref1,ref2,ref3,ref4,ref5,ref6}. The major cosmological constant problems are: fine tuning and cosmic coincidence which yet to be solve. Therefore, one may argue that the cosmological constant problem is the crisis of fundamental physics \cite{ref7,ref8}. It enforce cosmologist to think about the alternative of cosmological constant. This is why, in the literature various dark energy models with dynamical equation of state (EoS) parameter $\omega^{(eff)}$ have been studied (for example see \cite{ref9,ref10,ref11,ref12}), where $\omega^{(eff)} = -1$ represents the $\Lambda$CDM universe \cite{ref13,ref14}. Some other proposals to avoid these conceptual problems associated with $\Lambda$ have been reported from time to time \cite{ref15,ref16,ref17,ref18,ref19}. However, to concur with high precision observational data, we have to have small deviation from $\Lambda$CDM model or modification to general theory of relativity but indeed, we still do not have a promising and concrete fundamental theory to handle this issue \cite{ref20}. In 1961, Brans-Dicke \cite{ref21} had proposed a scalar-tensor theory in which the average expansion rate is modified due to alignment of scalar field with geometry while the geometrization of tensor field remains alone \cite{ref22}. So, both the scalar and tensor field have more or less intrinsic geometrical consequences and finally executes a more general method of geometrizing gravitation. Other word, despite of Einstein's general theory of relativity (GR), Brans-Dicke (BD) cosmology is not a fully geometrical theory of gravity. Brans-Dicke theory contains an additional parameter, the Brans-Dicke coupling $\omega$, as compared to
GR. It should be noted that while large $\omega$ refers to as important contribution of the Ricci scalar, small $\omega$ indicates the main contribution of the scalar field. Since Einstein's GR theory is recovered in the limit $\omega\to\infty$, BD theory could be considered as a good scale to quantify the accuracy of the predictions of GR against observational tests (reader is advised to see the textbooks, \cite{ref23,ref24}). The value of $\omega$ could be estimated from the astronomical and astrophysical observational data. Using solar system data obtained from Cassini-Huygens mission, it is estimated $\omega>40000$ at $2\sigma$ confident level (CL) \cite{ref25,ref26, ref27}. Nevertheless, constraining $\omega$ using cosmological data obtained from WMAP and Planck missions gave lower values for this parameter. Wu and Chen combined cosmic microwave background (CMB) data from 5 years of WMAP, and LSS measurements from the Sloan Digital Sky Survey (SDSS) release 4 \cite{ref28} and obtained $\omega > 97.8$ at $2\sigma$ CL. By using structure formation constraints, Acquaviva et al \cite{ref29} have obtained $\omega>120$ at $2\sigma$ CL. Li et al \cite{ref30} have used CMB temperature data from the Planck satellite, the 9 year polarization data from the WMAP and some others cosmological observations and found that $\omega$ varies in region $-407.0<\omega<175.87$ at $2\sigma$ CL. Hrycyna et al \cite{ref31} used supernovae type Ia and other cosmological data and found $\omega=\{-0.8606^{+0.8281}_{-0.1341},~ -1.1103^{+0.1872}_{-0.1.729},~ -2.3837^{+0.4588}_{-4.5459}\}$ for linear, oscillatory, and transient approaches to the de Sitter state respectively. Avilez and Skordis \cite{ref32} have reported strong cosmological constraints on coupling parameter of Brans-Dicke theory as $\omega > 692$ by using CMB data. We recommend readers to see Refs \cite{ref33,ref34,ref35} for theoretical bounds on the Brans-Dicke coupling constant.\\

Generally, in the case of zero space curvature $(k =0)$ and zero pressure $(p =0)$ it is possible to find analytical solution for BD equations in the absence of the cosmological constant $\Lambda$ \cite{ref36}. However, since original BD model does not have accelerating expansion phase, it is convenient to extend this model by adding a $\Lambda$-term as an effective contribution rather than a potential (note that the precise shape of the scalar field potential is not known yet \cite{ref37}). Some interesting BD solutions with a non-vanishing cosmological constant (henceforth $\Lambda$BD) could be found in Refs \cite{ref38,ref39,ref40,ref41,ref42,ref43}. In 1973 Gurevich et al \cite{ref44} obtained an exact analytical solution for pure BD with negative coupling constant, $\omega<0$, with no initial singularity. Since there is no cosmic acceleration in Gurevich et al solution, Tretyakova et al \cite{ref45} added a cosmological constant in order to extended Gurevich et al solution with negative $\omega$. They show that the scale factor may not vanish, unlike in the standard $\Lambda$CDM case. In 1984, Singh and Singh \cite{ref46} had investigated Brans-Dike cosmological model by choosing $\Lambda = \Lambda(\phi)$ and this idea was extended by Azad and Islam \cite{ref47} in an-isotropic and homogeneous Bianchi type I spacetime. A higher dimensional interacting scalar field and Higgs model is obtained by Qiang et al \cite{ref48}. Later on, using classical approach, Smolyakov \cite{ref49} has investigated Brans-Dike cosmological model with with effective value of $\Lambda$ in 5 D. Das and Banerjee \cite{ref50} have investigated model of accelerating universe with variable deceleration parameter in BD theory. In 2010, Setare and Jamil \cite{ref51} have constructed the chameleon model of holographic dark energy in BD theory and compared their results with those obtained via GR for large value of coupling constant $\omega$. In this connection, Yadav \cite{ref52} and Ali et al \cite{ref53} have investigated some non-isotropic, non-flat and in-homogeneous models of accelerating universe in BD scalar-tensor theory of gravitation. In this paper we obtain a new exact analytical solution for $\Lambda$BD model and use recent observational data namely CMB, BAO, and observational Hubble data (OHD) obtained from cosmic chronometric (CC) technique to constrain model parameters. It is worth mentioning that since in BD theory gravitational coupling $G$ is not a constant parameter, hence it's variation means that supernovae can no longer be considered as standard candles (see Ref \cite{ref54} for more details). We use Metropolis-Hastings algorithm to generate Markov chain Monte Carlo (MCMC) chains and estimate model parameters. Although the main goal of this paper is to constrain $\omega$ (as we show in the next section, instead of BD coupling constant we introduce new density parameter i.e $\Omega_{\phi}$ and constrain this parameter) and $G$, but, we also discuses other important quantities such as deceleration and Hubble parameters. The plan of this paper is as follows: Section \ref{sec:2} deals with the model and basic equations. We summary the computational technique used to fit model parameters to data by a numerical MCMC analysis in Section\ref{sec:3}. Section~ \ref{sec:4} deals with the results of our fits to data. Finally, we summarize our findings in Section~ \ref{sec:5}.
\section{Theoretical model and Basic equations}
\label{sec:2}
The Einstein's field equations in Brans-Dicke theory with cosmological constant are given by
\[
R_{ij}-\frac{1}{2}Rg_{ij}+\Lambda g_{ij}=\frac{8\pi}{\phi c^{2}}T_{ij}
\]
\begin{equation}
\label{BD-1}
-\frac{\omega}{\phi^{2}}\left(\phi_{i}\phi_{j}-\frac{1}{2}g_{ij}\phi_{k}\phi^{k}\right)-\frac{1}{\phi}(\phi_{ij}-g_{ij}\square\phi),
\end{equation}
and
\begin{equation}
\label{BD-2}
(2\omega+3)\square\phi=\frac{8\pi T}{c^{4}}+2\Lambda \phi,
\end{equation}
where $\omega$ is the Brans-Dicke coupling constant; $\phi$ is Brans-Dicke scalar field and $\Lambda$ is the cosmological constant.\\ 
The FRW space-time is read as
\begin{equation}
\label{FRW}
ds^{2} = c^{2}dt^{2}-a^{2}(t)(dx^{2}+dy^{2}+dz^{2}),
\end{equation}
where $a(t)$ is scale factor.\\
The energy momentum tensor of perfect fluid is given by
\begin{equation}
\label{em}
T_{ij}=(p+\rho)u_{i}u^{j}-pg_{ij}.
\end{equation}
Here, $p$ and $\rho$ are the isotropic pressure and energy density of the matter respectively and $u^{i}u_{i} =1$; where $u^{i}$ is the four velocity vector.\\
The field equations (\ref{BD-1}) for space-time (\ref{FRW}) are read as
\begin{equation}
\label{ef-1}
2\frac{\ddot{a}}{a}+\frac{\dot{a}^{2}}{a^{2}}+\frac{\omega \dot{\phi}^{2}}{2\phi^2}+2\frac{\dot{\phi}\dot{a}}{\phi a}+\frac{\ddot{\phi}}{\phi}=-\frac{8\pi}{\phi c^{2}}p+\Lambda c^{2},
\end{equation} 
\begin{equation}
\label{ef-2}
\frac{\dot{a}^2}{a^2}+\frac{\phi \dot{a}}{\phi a}-\frac{\omega \dot{\phi}^{2}}{6\phi^2}=\frac{8\pi}{3\phi c^2}\rho +\frac{\Lambda c^2}{3},
\end{equation}
\begin{equation}
\label{ef-3}
\frac{\ddot{\phi}}{\phi}+3\frac{\dot{\phi}\dot{a}}{\phi a} = \frac{8\pi(\rho-3p)}{(2\omega+3)c^{2}\phi}+\frac{2\Lambda c^{2}}{2\omega +3}.
\end{equation}
In this paper we use over dot to show derivatives with respect to time $t$.\\

Equations (\ref{ef-1}), (\ref{ef-2}) and (\ref{ef-3}) lead the following equation
\begin{equation}
\label{ef-4}
\Lambda c^{2} = 3\frac{\ddot{a}}{a}+3\frac{\dot{a}^{2}}{a^{2}}-\omega\frac{\ddot{\phi}}{\phi}-3\omega\frac{\dot{\phi\dot{a}}}{\phi a}+\frac{\omega \dot{\phi}^{2}}{\phi^{2}}.
\end{equation}
Before solving above questions we define the matter and cosmological constant density parameters as
\begin{equation}
\label{dp-1}
\Omega_{m} = \frac{8\pi \rho_{m}}{3c^2 H^2 \phi},~~\Omega_{\Lambda}=\frac{\Lambda c^2}{3H^2}, 
\end{equation}
where $\rho_{m} = \left(\rho_{m}\right)_{0}a^{-3}$ and $\{\Omega_{m}, \Omega_{\Lambda}\}$ stand for matter and $\Lambda$-term density parameters respectively. We also define the deceleration parameter (DP) $q$ and the scalar field deceleration parameter $q_{\phi}$ as bellow
\begin{equation}
\label{dp-1a}
q = -\frac{\ddot{a}}{aH^{2}} ~~~ q_{\phi} = -\frac{\ddot{\phi}}{\phi H^{2}}.
\end{equation}
Now, dividing equations (\ref{ef-2}) by $H^{2}$ and then using equation (\ref{dp-1}), we obtain

\begin{equation}
\label{n-1}
\Omega_{m}+\Omega_{\Lambda} = 1+\psi-\frac{\omega}{6}\psi^{2},
\end{equation}
where $\psi = \frac{\dot{\phi}}{\phi H}$. It should be noted that in BD gravity theory one can not use the usual relation between Hubble parameter and density in order to define density parameters, meaning that for a spatially-flat model the sum of density parameters don't quite sum to one.\\

We assume that the present universe is filled with pressure-less matter i. e. $p = 0$. So, putting $p = 0$ in equations (\ref{ef-1}), (\ref{ef-3}) and (\ref{ef-4}) then dividing by $3H^{2}$ and using equations (\ref{dp-1}) and (\ref{dp-1a}), we obtain the following relations
\begin{equation}
\label{n-2}
\Omega_{\Lambda} = \frac{\omega+3}{2\omega}\Omega_{m}-\frac{2\omega+3}{2\omega}q+\frac{2\omega+3}{6}\psi^{2}-\frac{2\omega+3}{2\omega}\psi,
\end{equation} 
\begin{equation}
\label{n-3}
1-\frac{3(\omega+1)}{2\omega}\Omega_{m}+\frac{2\omega+3}{2\omega}q+\frac{4\omega+3}{2\omega}\psi-\frac{\omega+1}{2}\psi^{2}=0,
\end{equation}
\begin{equation}
\label{n-4}
\Omega_{m} = q-\frac{\omega}{3}q_{\phi}+(\omega+1)\psi-\frac{\omega}{3}\psi^{2}.
\end{equation} 
Equations (\ref{n-3}) and (\ref{n-4}) lead to
\begin{equation}
\label{n-5}
q-(\omega+1)q_{\phi}+(3\omega+2)\psi = 2.
\end{equation}
The first integral of equation (\ref{n-5}) is read as
\begin{equation}
\label{n-6}
(\omega+1)\frac{\dot{\phi}}{\phi}-\frac{\dot{a}}{a} = \frac{\kappa}{\phi a^{3}},
\end{equation}
where $\kappa$ is the constant of integration.\\

The solution of equation (\ref{n-6}) has singularity at $a = 0$ and $\phi = 0$ which in turn gives $\kappa =0$. Thus equation (\ref{n-6}) could be written as
\begin{equation}
\label{n-7}
(\omega+1)\frac{\dot{\phi}}{\phi}-\frac{\dot{a}}{a} =0.
\end{equation}

Integrating equation (\ref{n-7}), finally we obtain
\begin{equation}
\label{n-8}
\phi = \left(\frac{a}{a_{0}}\right)^{\frac{1}{\omega+1}}.
\end{equation}
Dividing equation (\ref{n-7}) by $H$ and then integrating, we obtain\\

$\psi = \frac{1}{\omega+1}$.\\

Thus, equation (\ref{n-1}) leads to
\begin{equation}
\label{FRW-1}
\Omega_{m}+\Omega_{\Lambda} = 1+\frac{5\omega+6}{6(\omega+1)^{2}}.
\end{equation}
If we define the density of scalar field $\phi$ as
\begin{equation}
\label{Sf-1}
\Omega_{\phi}=-\frac{5\omega+6}{6(\omega+1)^{2}},
\end{equation}
then eq (\ref{FRW-1}) is written as
\begin{equation}
\label{FRW-2}
\Omega_{m}+\Omega_{\Lambda}+\Omega_{\phi} = 1.
\end{equation}

The scale factor $a$ and $\phi$ in connection with redshift $z$ are read as
\begin{equation}
\label{a-z}
a = \frac{a_{0}}{1+z},~~ \phi = \frac{1}{(1+z)^{\frac{1}{1+\omega}}}=\frac{1}{(1+z)^{2.5(\sqrt{1+0.96\Omega_{\phi}}-1)}},
\end{equation} 
where $a_{0}$ is the present value of scale factor.\\

Equations (\ref{dp-1}), (\ref{FRW-2}) and (\ref{a-z}) leads to 
\begin{equation}
\label{H-BD}
H_{BD}= \frac{H_{0}}{(1-\Omega_{\phi})^\frac{1}{2}}\left[\Omega_{m}(1+z)^{3+2.5(\sqrt{1+0.96\Omega_{\phi}}-1)}+\Omega_{\Lambda}\right]^\frac{1}{2}, 
\end{equation}
where $H_{0}$ is the present value of Hubble's parameter. Above equation clearly imposes a lower bound on the scalar field density as $\Omega_{\phi}\geq-1.04$. It is also interesting to note that, now, $\Omega_{\phi}\to 0$ is equivalent to $\omega \to \infty$ for which GR recovers from BD gravity theory. Obviously, as $\Omega_{\phi}\to 0$, $\Lambda$CDM model recovers from eq (\ref{H-BD})
\begin{equation}
\label{H}
H_{\Lambda CDM}=H_{0}[\Omega_{m}(1+z)^{3}+\Omega_{\Lambda}]^{\frac{1}{2}}.
\end{equation} 
We also note that a de-Sitter solution arises whenever
\begin{equation}
\label{de-sit}
q=-1\Rightarrow\dot{H}=0\Rightarrow H=H_{0}\Rightarrow a(t)\propto e^{H_{0}t}.
\end{equation} 
Since for our $\Lambda$BD model we have
\begin{equation}
\label{de}
q_{z=0}=\frac{\Omega_m \left(\sqrt{24 \Omega _{\phi }+25}-3\right) -4 \Omega _{\Lambda }}{4 \left(\Omega _{\Lambda }+\Omega _m \right)},
\end{equation}
it is clear to conclude that this condition can not be achieved in our $\Lambda$BD model (see \ref{apen-A} for derivation of DP an transition redshift).
\\
In the next section we use joint combination of three independent observational data including OHD, BAO, CMB to constrain both $\Lambda$CDM and $\Lambda$BD models with following parameters space

\begin{equation}
\label{GR-Sp}
{\bf\Theta_{\Lambda CDM}}= \{H_{0}, \Omega_m, \Omega_{\Lambda}, \Omega_{b}h^{2}, t(Gyr), q_{0}\},
\end{equation}
and
\begin{equation}
\label{BD-Sp}
{\bf\Theta_{\Lambda BD}}= \{H_{0}, \Omega_m, \Omega_{\Lambda}, \Omega_{b}h^{2}, t(Gyr), q_{0}, \Omega_{\phi}, \dot{G}/G\}.
\end{equation}
Note that $\Lambda$CDM model has only two free parameters i.e $\{H_{0}, \Omega_{m}\}$ while $\Lambda$BD model has three free parameters namely $\{H_{0}, \Omega_{m}, \Omega_{\phi}\}$, hence, all other parameters are derived parameters. 
\section{Data and Method}
\label{sec:3}
We use Metropolis-Hasting algorithm from the Pymc3 python package to generate MCMC chains for parameter spaces (\ref{GR-Sp}) and (\ref{BD-Sp}). For each parameter we run 4 parallel chains with 100000 iterations to stabilize the estimations. In order to confirm the convergence of the MCMC chains we perform well known Gelman-Rubin and Geweke tests. Moreover, we monitor the trace plots for good mixing and stationarity of the posterior distributions to confirm the convergence of all chains.
In our Bayesian analysis, we assume the following uniform priors for free parameters:
\begin{equation}
\label{eq32} H_{0} \sim U(60-800)~~~\Omega_{m}\sim U(0-0.5)~~~\Omega_{\phi} \sim U(-0.5-0.5).
\end{equation}
In bellow we briefly describe the cosmological data we have been used to constrain parameter spaces (\ref{GR-Sp}) and (\ref{BD-Sp}).\\
{\bf Observational Hubble Data}: we use $31H(z)$ datapoints (Table.~\ref{tab:1}) in the redshift range $.07\leq z\leq 1.965$ taken from Table 2 of Ref \cite{ref55}. The cosmic chronometric (CC) technique is used to determined these uncorrelated data. There reason why we use this data is behind the fact that OHD data obtained from CC technique is model-independent. In fact, the most massive and passively evolving galaxies based on the \textquotedblleft galaxy differential age\textquotedblright method is used to determine the CC data (see ref\cite{ref56} for more details). Since in this compilation all data are uncorrelated, we consider ${\bf C_{ij}}=\mbox{diag}(\sigma_{i}^{2})$ as covariance matrix of this class of data. The chisqure for this data is given by
\begin{equation}
\label{chi-H} \chi^{2}_{HOD}= (H(z)-H_{0})^{T}{\bf C}^{-1}(H(z)-H_{0}),
\end{equation}
\begin{table}[ht]
\caption{Hubble parameter versus redshift data.}
\centering
\setlength{\tabcolsep}{25pt}
\scalebox{0.7}{
\begin{tabular} {cccc}
\hline
\hline
$H(z)$    &  $\sigma_{H}$   &  $z$  & Reference\\[0.5ex]
			
\hline{\smallskip}
69 &      19.6     & 0.070   & \cite{ref57}\\
			
69 &      12       & 0.090   & \cite{ref58} \\
			
68.6 &    26.2     & 0.120   & \cite{ref57}\\
			
83 &      8        & 0.170   & \cite{ref58} \\
			
75 &      4        & 0.179   & \cite{ref59} \\
			
75 &      5        & 0.199   & \cite{ref59} \\
			
72.9 &    29.6     & 0.200   & \cite{ref57} \\
			
77 &      14       & 0.270   & \cite{ref58} \\
			
88.8 &    36.6     & 0.280   & \cite{ref57} \\ 
			
83  &     14       &0.352    & \cite{ref59} \\

83  &     13.5     &0.3802   & \cite{ref60} \\
			
95  &     17       &0.400    & \cite{ref58} \\
			
77  &     10.2     &0.4004   & \cite{ref60} \\
			
87.1  &   11.2     &0.4247   & \cite{ref60} \\
			
92.8  &   12.9     &0.4497   & \cite{ref60} \\
			
89    &   50       &0.47     & \cite{ref61} \\
			
80.9  &   9        &0.4783    & \cite{ref60} \\
			
97 &      62       & 0.480   &\cite{ref62}  \\
						
104 &     13       &0.593    & \cite{ref59} \\
		
92 &      8        &0.680    &\cite{ref59}  \\
			
105 &     12       &0.781    & \cite{ref59} \\
			
125 &     17       &0.875    & \cite{ref59} \\
			
90 &      40       &0.880    & \cite{ref62} \\
			
117 &     23       &0.900    & \cite{ref58} \\ 
			
154 &     20       &1.037    & \cite{ref59} \\
			
168 &     17       & 1.300   & \cite{ref58} \\ 
			
160  &    33.6     &1.363    & \cite{ref63} \\
			
177 &     18       &1.430    & \cite{ref58} \\
			
140 &     14       &1.530    & \cite{ref58} \\
			
202 &     40       &1.750    & \cite{ref58} \\
			
186.5 &   50.4     &1.965    & \cite{ref63} \\
			
\hline
\hline
\end{tabular}}
\label{tab:1}
\end{table}
{\bf CMB}: It is shown \cite{ref64} that the existence of dark energy could affect CMB power spectrum in at least two ways. First, the change of the angular diameter distance causes a shift in positions of CMB acoustic peaks from the last scattering surface to today. Second, the presence of dark energy results in fluctuation of gravitational potentials which in turn leads to the late-time integrated Sachs-Wolfe effect. As the first effect is more important to constrain dark energy models, we only use CMB distance measurements in our statistical analysis. In this regard, we consider following two CMB shift parameters (for flat space-time).
\begin{equation}
\label{shift-R} R=\sqrt{\Omega_{m}}\int_{0}^{z_\ast}\frac{dz}{E(z)},
\end{equation}
and 
\begin{equation}
\label{shift-l} l_{a}=\frac{\pi r(z_{\ast})}{r_{s}(z_{\ast})},
\end{equation}
while the average acoustic structure is determined by the acoustic scale $l_{a}$, the shift parameter $R$ is associated with
the overall amplitude of CMB acoustic peak. In eq (\ref{shift-l}) the comoving distance to the CMB decoupling surface $r(z_{\ast} )$ and the comoving sound horizon at the CMB decoupling $r_{s} (z_{\ast} )$ are given by
\begin{equation}
\label{shift-r} r(z_{\ast} )=\int_{0}^{z_{\ast}}\frac{dz}{H(z)}, ~~ r_{s} (z_{\ast}) =\frac{1}{H_{0}}\int_{z}^{\infty}\frac{dz'c_{s}(z')}{E(z')},
\end{equation}
where $E(z)^{2}=H(z)/H_{0}$, and the sound speed squared $c_{s}$ of baryon fluid coupled with photons ($\gamma$) is given by \cite{ref65}
\begin{equation}
\label{S-S} c_{s}=\frac{1}{\sqrt{3[1+R_{b}/(1+z)]}},~~ R_{b}=31500 \Omega_{b}h^{2}\left(\frac{2.7255}{2.7}\right)^{-4},
\end{equation}
where $h$ is the normalized Hubble constant and $z_{\ast}=1090$ is decoupling redshift. We use the CMB distance priors based and Planck 2015 collaboration \cite{ref3} to constrain our models. According to Planck 2015 data, the mean values of shift and acoustic scale parameters are $\langle R\rangle=1.7488$ and $\langle l_{a}\rangle=301.76$ with the deviations $\sigma(R)=0.0074$ and $\sigma(l_{a})=0.14$ respectively. For this data, the inverse of the covariance matrix is given by
\begin{equation}
\label{co-m} 
\mathbf{C}^{-1} =\begin{pmatrix}
1.412 & -0.762\\[0.2ex]
-0.762& 1.412
\end{pmatrix}.
\end{equation}
Finally, The chisquare associated with the CMB data could be obtained as
\[
 \chi_{CMB}^{2}= (l_{a}-301.76)^{2} \times 1.412 + (R-1.7488)^{2}\times 1.412 
\]
\begin{equation}
\label{chi-cmb}+2(l_{a}-301.76)(R-1.7488)\times (-0.762).
\end{equation}
{\bf BAO}: The counteracting forces of pressure and gravity result in periodic fluctuations of the density
of baryonic matter which represent by BAO. We use almost the same BAO data points were used in the
WMAP 9-year analysis \cite{ref66}. This includes ten numbers of $r_{s} (z_{d} )/D_{V}(z)$ extracted from the 6dFGS \cite{ref67}, SDSS-MGS \cite{ref68}, BOSS \cite{ref69}, BOSS CMASS \cite{ref70}, and WiggleZ \cite{ref71}
surveys. Here $D_{V}(z)$ which is given by
\begin{equation}
\label{Dv}
D_{V}(z)=\left[r^{2}(z)\frac{cz}{H(z)}\right]^{\frac{1}{3}},
\end{equation}
is the effective distance measure related to the BAO scale \cite{ref72}, $c$ is the speed of light, and $r_{s}(z_{d})$ is the comoving sound horizon size at the drag epoch. The drag redshift $z_{d}$,  which is the redshift at which baryons are released from photons has the following fitting formula \cite{ref73}
\begin{equation}
\label{zd}
z_{d}=\frac{1291(\Omega_{m}h^{2})^{0.251}}{1+0.659(\Omega_{m}h^{2})^{0.828}}[1+b_{1}(\Omega_{b}h^{2})^{b_{2}}],
\end{equation}
where
\begin{equation}
\label{b1}
b_{1}=0.313(\Omega_{m}h^{2})^{-0.419}[1+0.607(\Omega_{m}h^{2})^{0.674}],
\end{equation}
and
\begin{equation}
\label{b2}
b_{2}=0.238(\Omega_{m}h^{2})^{0.223}.
\end{equation}
The chisquare statistics for BAO could be written as 
\begin{widetext}
\begin{multline}
\chi^{2}_{BAO}=\\\frac{1}{0.015^{2}}\left[\frac{r_{s}(z_{d})}{D_{V}(z=0.106)}-0.336\right]^{2}
+\frac{1}{(\frac{25}{149.69})^{2}}\left[\frac{D_{V}(z=0.15)}{r_{s}(z_{d})}-\frac{664}{148.69}\right]^{2} 
+\frac{1}{(\frac{25}{149.28})^{2}}\left[\frac{D_{V}(z=0.32)}{r_{s}(z_{d})}-\frac{1264}{149.28}\right]^{2}\\
+\frac{1}{(\frac{16}{147.78})^{2}}\left[\frac{D_{V}(z=0.38)}{r_{s}(z_{d})}-\frac{1477}{147.78}\right]^{2}
+\frac{1}{0.0071^{2}}\left[\frac{r_{s}(z_{d})}{D_{V}(z=0.44)}-0.0916\right]^{2}
+\frac{1}{(\frac{19}{147.78})^{2}}\left[\frac{D_{V}(z=0.51)}{r_{s}(z_{d})}-\frac{1877}{147.78}\right]^{2}\\
+\frac{1}{(\frac{20}{149.28})^{2}}\left[\frac{D_{V}(z=0.57)}{r_{s}(z_{d})}-\frac{2056}{149.28}\right]^{2}
+\frac{1}{0.0034^{2}}\left[\frac{r_{s}(z_{d})}{D_{V}(z=0.6)}-0.0726\right]^{2}
+\frac{1}{(\frac{22}{147.78})^{2}}\left[\frac{D_{V}(z=0.61)}{r_{s}(z_{d})}-\frac{2140}{147.78}\right]^{2}\\
+\frac{1}{0.0032^{2}}\left[\frac{r_{s}(z_{d})}{D_{V}(z=0.73)}-0.0592\right]^{2}
\end{multline}
\end{widetext}
Finally, since these three datasets are independent, the total chisqure could be given by $\chi^{2}_{tot}=\chi^{2}_{OHD}+\chi^{2}_{CMB}+\chi^{2}_{BAO}$. Therefore, we evaluated the following total likelihood for statistical analysis. 
\begin{equation}
\label{eqlik} \mathcal{L}_{tot}\propto \exp\left(-\frac{1}{2} \chi^{2}_{tot}\right).
\end{equation}
\section{Results}
\label{sec:4}
\begin{figure}[h!]
\centering
\includegraphics[width=8.5cm,height=8cm,angle=0]{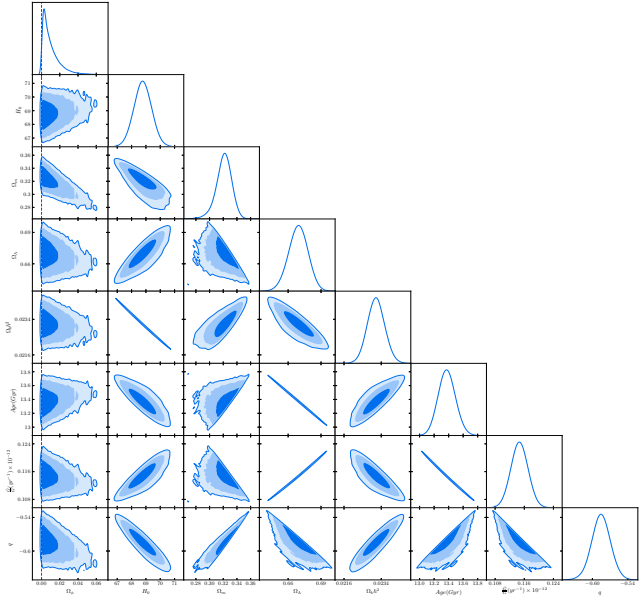}
\caption{One-dimensional marginalized distribution, and three-dimensional contours with $68\%$ CL, $95\%$ CL, and $99\%$ CL for  parameter space {\bf$\Theta_{\Lambda DB}$} using {\bf CC+ABO+CMB} data. The vertical dashed red line stands for $\Omega_{\phi}=0$.}
\label{fig1}
\end{figure}

\begin{figure}[h!]
\includegraphics[width=8.5cm,height=8cm,angle=0]{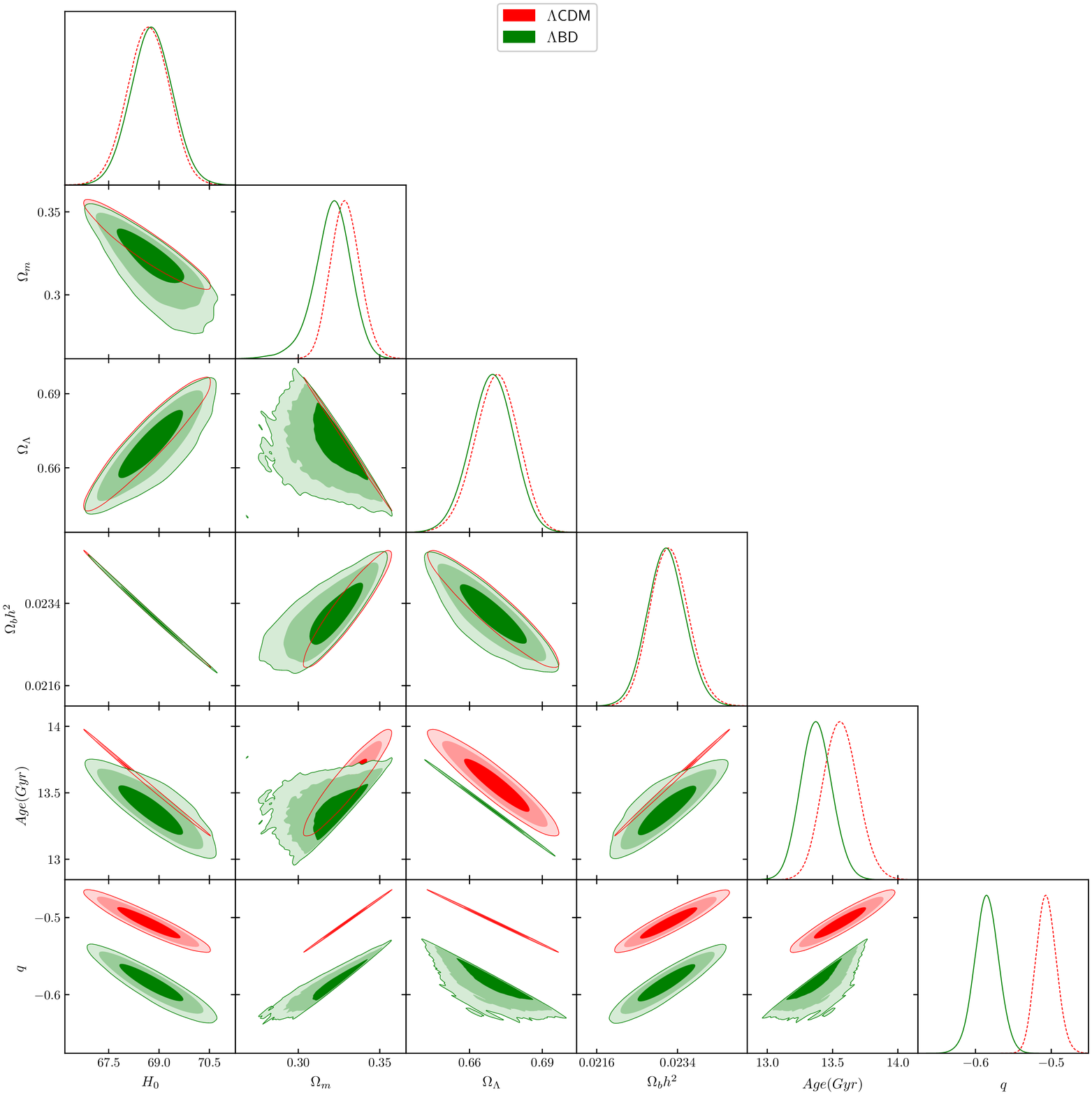}
\caption{One-dimensional marginalized distribution, and two-dimensional contours with $68\%$ CL, $95\%$, and $99\%$ CL for flat $\Lambda$CDM and $\Lambda$BD models fitted over {\bf CC+ABO+CMB} data.}
\label{fig2}
\end{figure}

\begin{figure}[h!]
\includegraphics[width=8.5cm,height=8cm,angle=0]{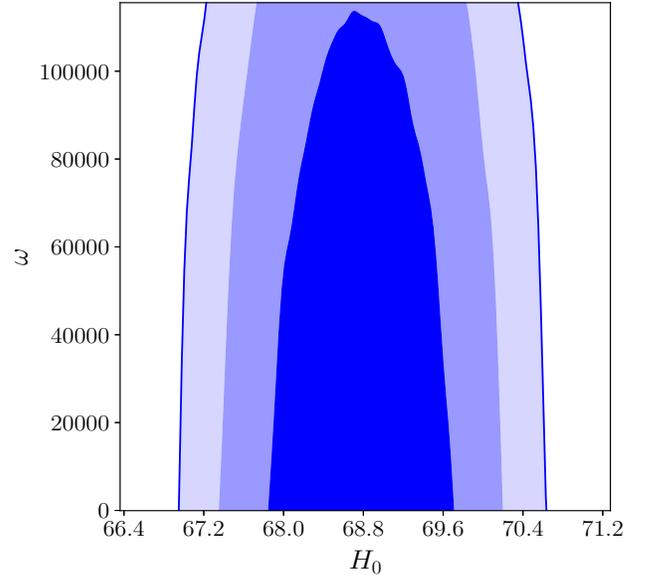}
\caption{Constraints in the $(\omega,H_{0})$ plane with $1\sigma$-$3\sigma$ confident level in the flat $\Lambda$BD model fitted over {\bf CC+ABO+CMB} data.}
\label{fig3}
\end{figure}

We have listed our statistical analysis on parameter spaces (\ref{GR-Sp}) \& (\ref{BD-Sp}) using the joint combination of OHD+CMB+BAO at 1$\sigma$ and $2\sigma$ error in Table.~\ref{tab:2}. Also, Figures.~\ref{fig1} \& \ref{fig2} show the contour plots, at $1\sigma$, $2\sigma$, and $3\sigma$ confidence levels, for both $\Lambda$BD and $\Lambda$CDM models using joint combination of three datasests respectively. First we discuss the $\Lambda$BD model, which contain $\omega$ as an additional parameter to $\Lambda$CDM model. From Table.~\ref{tab:2} we observe that $\Omega_{\phi}$ is restricted in the interval $0.0001<\Omega_{\phi}<0.0127$ at $1\sigma$ CL with the best fit $\Omega_{\phi}=0.0027$, which in turn gives the best fit for BD constant parameter as $\omega=308.452$. Our analysis show that the BD parameter is constrained to $\omega>\{211,1560,8460\}$ at $68\%, ~95\%$, and $~99\%$ confidence level respectively. We have compared our result of $\omega$ to those obtained from other researches in Table.~\ref{tab:3}. From this table we observe that our estimated bound on $\omega$ is much stronger than those reported by previous works. While our computed $\omega$ is comparable with that obtained by Avilez and Skordis \cite{ref32} at $95\%$ CL, but our analysis put much stronger bound on this parameter at $99$ CL. Also at $68\%$ CL our obtained bound is comparable with what computed by Li et al\cite{ref30}. Figures.~\ref{fig3} shows the $1\sigma$-$3\sigma$ contour plot of $(\omega,H_{0})$ pair.

\begin{table}
\caption{Results from the fits of the flat $\Lambda$CDM and $\Lambda$BD models to the data at $2\sigma$ \& $2\sigma$ confidence levels.}
\centering
\setlength{\tabcolsep}{2pt}
\scalebox{0.7}{
\begin{tabular}{c p{1cm} cccc}
\hline
\hline
& & Parameter & $\%68$ & $\%95$ & Best Fit \\
\hline
\multirow{7}{*}{\bf{$\Lambda$CDM}} &
\multirow{1}{*}{Fit} &
			
$H_{0}$ &  $68.68\pm 0.62$ & $68.7^{+1.2}_{-1.2}$ &  $67.90$ \\[.2cm]
&  & $\Omega_{m}$ &  $0.3290\pm 0.0089$ & $0.329^{+0.018}_{-0.017}$ &  $0.344$ \\[.2cm]

\cline{2-6}
& \multirow{6}{*}{Derived} &
$\Omega_{\Lambda}$ & $0.6710\pm 0.0089$ & $0.671^{+0.017}_{-0.018}$ & $0.655$ \\[.2cm]
&&$\Omega_{b}h^{2}$ & $0.02322\pm 0.00042$ & $0.02322^{+0.00086}_{-0.00079}$ & $0.023$ \\[.2cm]
&& $Age(Gyr)$ & $13.56\pm 0.13$ & $13.56^{+0.27}_{-0.25}$ & $13.73$  \\[.2cm]
&&$q$ & $-0.507\pm 0.013$  & $-0.507^{+0.027}_{-0.025}$  & $-0.482$ \\[.2cm]
\hline
\hline
\multirow{10}{*}{\bf{$\Lambda$BD}} &
\multirow{1}{*}{Fit} &
$H_{0}$ &  $68.78\pm 0.62$ &    $68.8^{+1.2}_{-1.2}$ &  $69.334$ \\[.2cm]
& &$\Omega_{\phi}$ & $0.0101^{+0.0026}_{-0.010}$ & $0.010^{+0.021}_{-0.012}$ & $0.0027$ \\[.2cm]
& &$\Omega_{m}$ &  $0.321^{+0.012}_{-0.0096}$   & $0.321^{+0.023}_{-0.024}$ &  $0.318$ \\[.2cm]
\cline{2-6}
& \multirow{10}{*}{Derived} &
$\Omega_{\Lambda}$ & $0.6692\pm 0.0089$ & $0.669^{+0.017}_{-0.018}$ & $0.678$ \\[.2cm]
&&$\omega$ & $> 211$  & $> 1560$  & $308.452$ \\[.2cm]
&&$\Omega_{b}h^{2}$ & $0.02315\pm 0.00042$ & $0.02315^{+0.00084}_{-0.00081}$ & $0.0227$ \\[.2cm]
&& $Age(Gyr)$ & $13.38\pm 0.12$ & $13.38^{+0.24}_{-0.23}$ & $13.246$  \\[.2cm]
&& $\frac{\dot{G}}{G}(10^{-12}yr^{-1})$ & $0.1147\pm 0.0024$ & $0.1147^{+0.0046}_{-0.0046}$ & $0.117$  \\[.2cm]
&&$q$ & $-0.585\pm 0.016$  &  $-0.585^{+0.032}_{-0.032}$  & $-0.595$ \\[.2cm] 
\hline
\hline
\end{tabular}}
\label{tab:2}
\end{table}
\begin{table}[h!]
\caption{Constraints on the BD coupling constant from different researches.}
\centering
\scalebox{0.8}{
\begin{tabular} {ccccccc}
\hline
BD parameter   &   Method\\[0.5ex]
\hline
\hline{\smallskip}
$\omega>\{80(3\sigma), 120(2\sigma)\}$&CMB+WMAP+LSS &\cite{ref29}\\
			
$\omega>97.8(2\sigma)$ &WMAP5+SDSSLRG &\cite{ref74}\\
			
$\omega>\{337.34(1\sigma), 181.65(2\sigma)\}$ &WMAP9pol+BAO&\cite{ref30}  \\
			
$\omega>\{1834(2\sigma),890(3\sigma)\}$& PLANCKTEMP+WMAP9pol&\cite{ref32}\\
$\omega>\{211(1\sigma),1560(2\sigma), 8460(3\sigma) \}$& CC+CMB+BAO&This paper\\
[0.5ex]
\hline
\end{tabular}}
\label{tab:3}
\end{table}
\begin{figure}[h!]
\includegraphics[width=8.5cm,height=8cm,angle=0]{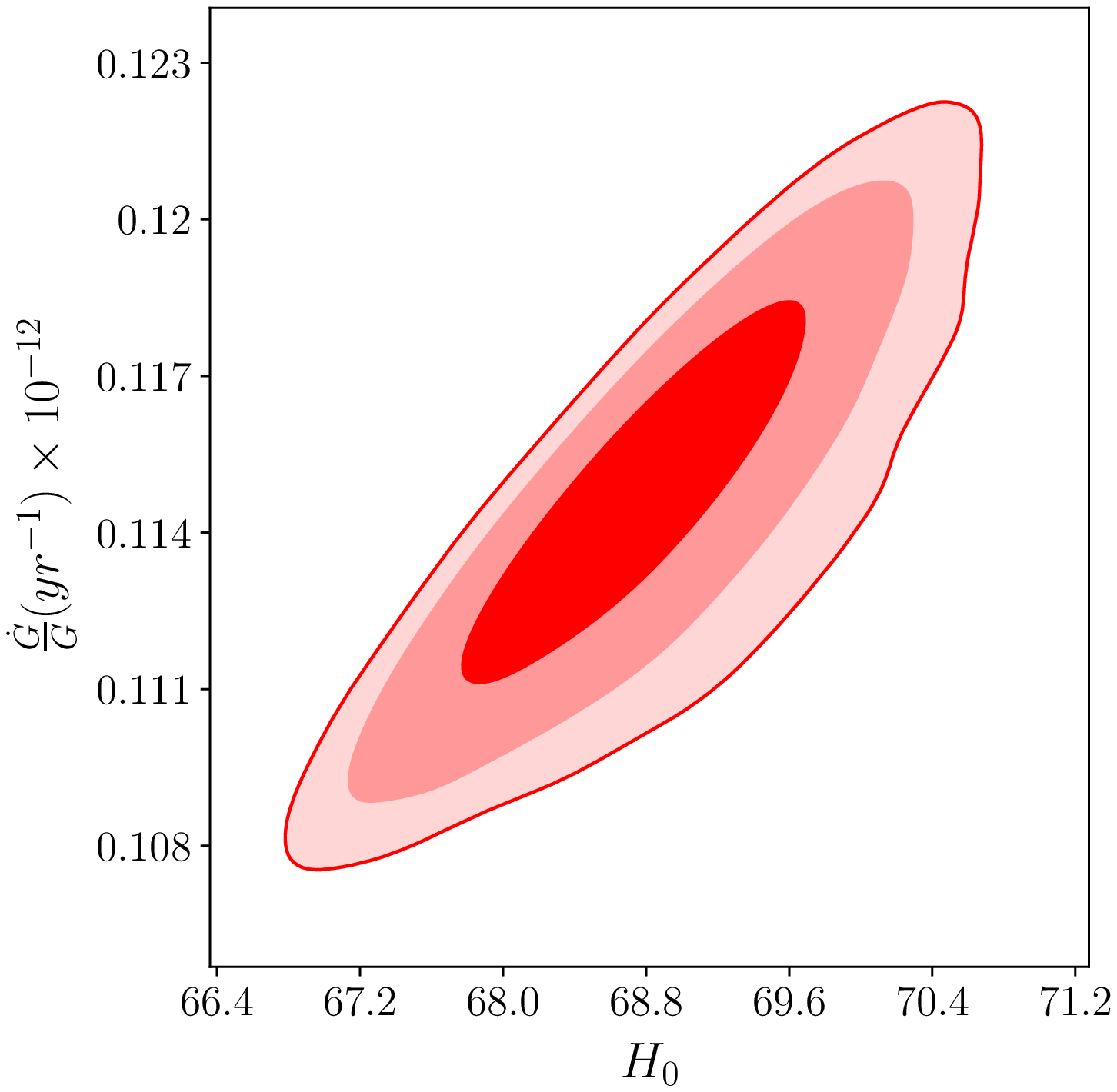}
\caption{Constraints in the $(\dot{G}/G,H_{0})$ plane with $1\sigma$-$3\sigma$ confident level.}
\label{fig4}
\end{figure}
\begin{figure}[h!]
\includegraphics[width=8.5cm,height=8cm,angle=0]{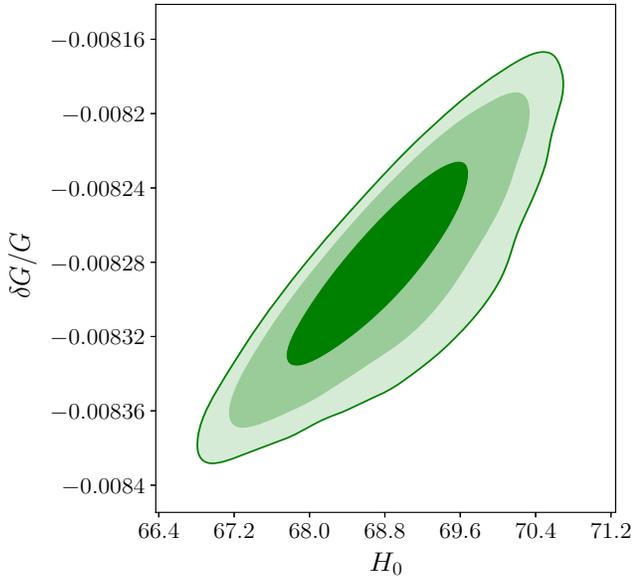}
\caption{Constraints in the $(\delta G/G,H_{0})$ plane with $1\sigma$-$3\sigma$ confident level.}
\label{fig5}
\end{figure}
We know that in the context of BD theory, the gravitational constant $G$ is also underwent evolution from recombination era to the current time \cite{ref73}. Therefor, one can estimate the variation in $G$ throughout variations of $\Omega_{\phi}~\mbox{or}~\omega$ as these two parameters are correlated. For this purpose, we inter two new derived parameters namely $\dot{G}/G\equiv -\dot{\phi}/\phi$ and $\delta G/G\equiv (G_{rec}-G_{0})/G_{0}$ which is the integrated change of gravitational constant since the epoch of recombination in our MCMC code. Figures.~\ref{fig4} \& \ref{fig5} depict 2-dimensional contours of marginalized likelihood distributions of $H_{0}$ versus $\dot{G}/G$ and $\delta G/G$ respectively. The $68$\% marginalized limits on these two parameters are as bellow.
\[
1.150\times 10^{-13}<\dot{G}/G<1.198\times 10^{-13},
\]
\begin{equation}
\label{marg}    -0.0084<\delta G/G<-0.0082,
\end{equation}
with the best-fit values
\begin{equation}
\label{best}    \dot{G}/G=1.17\times 10^{-13},~~\delta G/G=-0.00825.
\end{equation}
We have summarized constraints on $\dot{G}/G$ obtained from different methods in Table.~\ref{tab:4} which is an update of Table I of Ref \cite{ref30}. From this table we observe that our constrain is much tighter with respect to other listed previous constraints. 
\begin{table}[h!]
\caption{Constraints on the BD coupling constant from different researches.}
\centering
\scalebox{0.8}{
\begin{tabular} {cccccccccc}
\hline
$\dot{G}/G(10^{-13}yr^{-1})$   &   Method\\[0.5ex]
\hline
\hline{\smallskip}
$2\pm7$&Lunar laser ranging &\cite{ref75}\\
			
$0\pm4$ &Big bang nucleosynthesis &\cite{ref76,ref77}\\
			
$0\pm16$ &Helioseismology &\cite{ref78}  \\
			
$-6\pm20$& Neutron star mass&\cite{ref79}\\

$20\pm40$& Viking lander ranging&\cite{ref80}\\

$40\pm50$& Binary pulsar&\cite{ref81}\\

$-96\sim 81(2\sigma)$& CMB(WMAP3)&\cite{ref82}\\

$-17.5 \sim 10.5(2\sigma)$& WMAP5+SDSSLRG&\cite{ref74}\\

$-1.42^{+2.48+4.38}_{-2.27-4.74}(1\sigma 2\sigma)$& Planck+WP+BAO&\cite{ref30}\\

$1.174^{+0.0024+0.0046}_{-0.0024-0.0046}(1\sigma 2\sigma)$& CC+CMB+BAO&This paper\\
[0.5ex]
\hline
\end{tabular}}
\label{tab:4}
\end{table}
From Table.~\ref{tab:2} we see that the estimated Hubble constant $H_{0}$ for both $\Lambda$CDM and $\Lambda$BD models are in excellent agreement with those of Chen \& Ratra ($68 \pm 2.8$) \cite{ref83}, Aubourg et al (BAO: $67.3 \pm 1.1$) \cite{ref84}, Chen et al ($68.4^{+2.9}_{-3.3}$) \cite{ref85}, Aghanim et al (Planck 2018: $67.66 \pm 0.42$) \cite{ref4}, and 9-years WMAP mission ($68.92^{+0.94}_{-0.95}$) \cite{ref86}. Figure.~\ref{fig6} depicts the robustness of our fits for $H(z)$. When we fit both $(\Theta_{\Lambda CDM}, \Theta_{\Lambda BD})$ to CC+CMB+BAO data we obtain $(\Omega_{m}=0.329\pm0.0089, \Omega_{\Lambda}=0.671\pm0.0089)$ and $(\Omega_{m}=0.321^{+0.012}_{-0.0096}, \Omega_{\Lambda}=0.669\pm0.0089)$ for these models respectively. Obviously, these results are in excellent agreement with those obtained by Planck 2018 collaboration\cite{ref4} ($\Omega_{m}=0.3103 \pm 0.0057, \Omega_{\Lambda}=0.6897 \pm 0.0057$).
\begin{figure}[h!]
\includegraphics[width=8.5cm,height=6cm,angle=0]{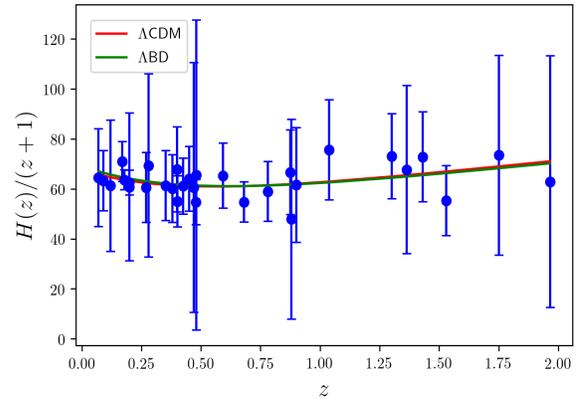}
\caption{The plot of Hubble rate  versus the redshift $z$. The points with bars indicate the experimental data summarized in Table.~\ref{tab:1}}
\label{fig6}
\end{figure}
\begin{figure}[h!]
\includegraphics[width=8.5cm,height=6cm,angle=0]{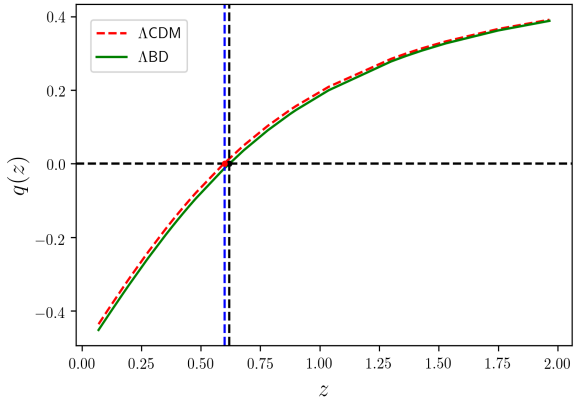}
\caption{The plot of deceleration parameter versus the redshift $z$ for Both $\Lambda$CDM \& $\Lambda$BD models. Filled circles show the best fit values of DP at transition red-shift $z_{t}$.}
\label{fig7}
\end{figure}
The computed values of deceleration parameter for $\Lambda$CDM \& $\Lambda$BD models are $q_{\Lambda CDM}=-0.507^{+0.013+0.027+0.07}_{-0.013-0.025-0.032}$ and $q_{\Lambda BD}=-0.585^{+0.016+0.032+0.042}_{-0.016-0.032-0.044}$ at $1\sigma-2\sigma$ respectively. Our estimated values of DP for both models are in good agreement with those reported in Refs \cite{ref87}, \cite{ref88}, \cite{ref89}, and \cite{ref90}. The dependence of deceleration parameter $q(z)$ has been plotted in Figures.~\ref{fig7} as a function of redshift $z$. Our computations show that both $\Lambda$BD \& $\Lambda$CDM models inters the accelerating expansion phase almost at the same time (also see fig.~\ref{fig7}). We also estimated transition redshift $z_{t}$ for both models as a derived parameter in our MCMC code. We found $z_{t}=0.602^{+0.021+0.041+0.054}_{-0.021-0.041-0.055}$ for 
$\Lambda$BD and $z_{t}=0.599^{+0.021+0.041+0.055}_{-0.021-0.041-0.054}$ for $\Lambda$CDM models at $1\sigma-3\sigma$ CL.
Figure.~\ref{fig8} shows the $1\sigma-3\sigma$ contour plot of ($H_{0}-z_{t}$) pair. It has recently been shown \cite{ref91} that the transition redshift should be restricted in a specific interval as $0.33<z_{t}<1$. For Both models, the computed $z_{t}$ is in good agreement with those obtained in Refs\cite{ref92,ref93,ref94}
\begin{figure}[h!]
\includegraphics[width=8.5cm,height=6cm,angle=0]{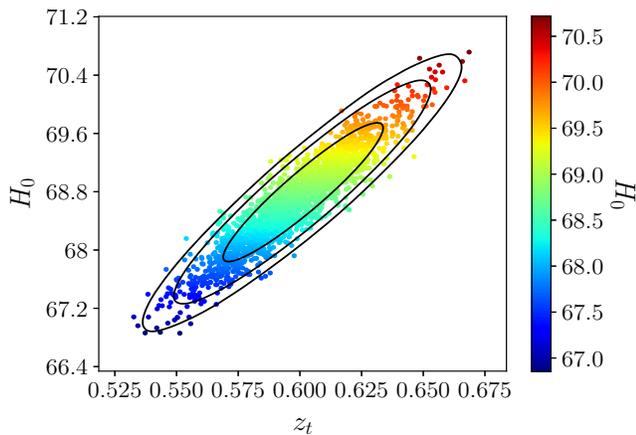}
\caption{$H_{0}-z_{t}$ plane with $1\sigma-3\sigma$ confident level. Solid contours stand for $\Lambda$BD model}
\label{fig8}
\end{figure}
\section{Concluding Remarks}
\label{sec:5}
In this paper first we tried to obtain a new exact solution for Brans-Dicke equations with the cosmological constant in the spatially flat Robertson-Walker metric. We defined new density parameter for scalar field i.e $\Omega_{\phi}$ which recovers GR theory when tends to zero (this is ,of course, equivalent to the case when $\omega\to \infty$). The latest observational data namely OHD (CC data), CMB, and BAO have been used to constrain parameter spaces \ref{BD-Sp} (Brans-Dicke theory) \& \ref{GR-Sp} (General relativity theory). To this aim, We have used Metropolis-Hasting algorithm to perform MCMC analysis. The marginalized bounds on BD coupling constant are obtained as $\omega>(211, 1560)$ at $68\%$ and $95\%$ confident levels respectively. Our estimations put tighten constrain on the Brans-Dicke model compared with previous works. Moreover, we defined two additional derived parameters in MCMC code namely the rate of change of the gravitational constant $\dot{G}/G$ and the integrated change of this parameter $\delta G/G$ since the epoch of recombination to explore whether $G$ is a \textgravedbl constant\textacutedbl or not. The marginalized $1\sigma$ limits are given by equation \ref{marg}. These bounds are in excellent agreement with the precision of Solar System experiments. Generally, our computations do not show any significant deviation from general theory of relativity.
When we compare the distribution of other cosmological parameters of \ref{BD-Sp} \& \ref{GR-Sp} spaces, we find out that the introduction of Brans-Dicke gravity does not affect the best-fit values and estimated errors.


\section*{APPENDIX A}
\label{apen-A}
It is well known that at a specific redshift called transition, $z_{t}$ the expansion phase of universe changes from decelerating to accelerating. To obtain this special redshift, first we derive deceleration parameter which is defined as
\setcounter{equation}{0}
\renewcommand{\theequation}{A\arabic{equation}}
\begin{equation}
q(z)=-\frac{1}{H^{2}}\left(\frac{\ddot{a}}{a}\right)=\frac{(1+z)}{H(z)}\frac{dH(z)}{dz}-1.
\end{equation}
Using \ref{H-BD} in above equation, and after some algebra we obtain
\begin{equation}
q(z)=\frac{\Omega_{m}(\sqrt{24\Omega_{\phi}+25}-3)(1+z)^{\frac{1}{2}(\sqrt{24\Omega_{\phi}+25}+1)}}{4\left[\Omega_{\Lambda}+\Omega_{m}(1+z)^{\frac{1}{2}(\sqrt{24\Omega_{\phi}+25}+1)}\right]}.
\end{equation}
In the other hand, transition redshift could be defined by the condition $q(z_{t}) = \ddot{a}(z_{t}) = 0$. Applying this condition in above equation and after some algebra finally we get transition redshift as follows.
\begin{equation}
z_{t}=\left[\frac{4\Omega_{\Lambda}}{\Omega_{m}(\sqrt{24\Omega_{\phi}+25}-3)}\right]^{\frac{2}{\sqrt{24\Omega_{\phi}+25}+1}}-1.
\end{equation}
It is straightforward to show that in flat-$\Lambda$CDM model the transition redshif is given by
\begin{equation}
z_{t}=\left(\frac{2\Omega_{\Lambda}}{\Omega_{m}}\right)^\frac{1}{3}-1.
\end{equation}


\begin{thebibliography}{99}
	
\bibitem {ref1}
O. Akarsu et al. arXiv: 1903.06679v1 [gr-qc] (2019).
\bibitem {ref2}
E. Komastu et al. Astrophys. J. Suppl. Ser. \textbf{192}, 18 (2011).
\bibitem {ref3}
P. A. R. Ade et al. Astron. Astrophys. \textbf{594}, A13 (2016).
\bibitem {ref4}
N. Aghamin et al. [Plank Collaboration], arXiv: 1807.06209 [astro-ph.CO] (2018).
\bibitem {ref5}	
P. J. E. Peebles and B. Ratra, Rev. Mod. Phys. \textbf{75}, 559 (2003)
\bibitem {ref6}
T. Padmanabhan, Phys. Rept. \textbf{380}, 234 (2003)
\bibitem{ref7}  
S. Weinberg, Rev. Mod. Phys. \textbf{61}, 1 (1989)
\bibitem {ref8}
P.J.E. Peebles, B. Ratra, Rev. Mod. Phys. \textbf{75}, 559 (2003)
\bibitem{ref9}
V. Sahni, A. Shafieloo and A. A. Starobinsky, Astrophys. J. \textbf{793}, 2 (2014)
\bibitem{ref10} 
A. Shafieloo, B. L. Huillier and A. A. Starobinsky, Phys. Rev. D \textbf{98}, 083526 (2018)
\bibitem{ref11} 
A. K. Yadav, Astrophys. Space Sc. \textbf{335},565 (2011)
\bibitem{ref12} 
A. K. Yadav, Astrophys. Space Sc. \textbf{361},276 (2016)
\bibitem{ref13} 
G. K. Goswami, Research in Astron. Astrophys. \textbf{17}, 27 (2017); arXiv: 1710.0728 [gr-qc] (2017)
\bibitem{ref14} 
E. J. Copeland, M. Sami and S. Tsujikawa, Int. J. Mod. Phys. D \textbf{15}, 1753 (2007)
\bibitem{ref15} 
E.J. Copeland, A.R. Liddle and D. Wands, Phys. Rev. D \textbf{57}, 4686 (1988)
\bibitem{ref16} 
P.J.E. Peebles and B. Ratra, Astrophys. J. Lett. \textbf{325}, L17 (1988)
\bibitem{ref17} 
I. Zlatev, L. Wang and P.J. Steinhardt, Phys. Rev. Lett. \textbf{82}, 896 (1999)
\bibitem{ref18} 
N. Bartolo and M. Pietroni, Phys. Rev. D \textbf{61}, 023518 (1999)
\bibitem{ref19} 
E. J. Copeland, M. Sami and Shinnji Tsujikava, Int. J. Mod. Phys. D \textbf{15}, 1753 (2006)
\bibitem{ref20} 
R. R. Caldwell and M. Kamionkowski, Ann. Rev. Nucl. Part. Sci. \textbf{59}, 397 (2009)
\bibitem{ref21} 
C. Brans and R. H. Dicke, Phys. Rev. \textbf{124}, 925 (1961)
\bibitem{ref22} 
R. H. Dicke, ibid \textbf{125}, 2163 (1962)
\bibitem{ref23}
Y. Fujii, K.-I. Maeda, The Scalar-Tensor Theory of Gravitation (Cambridge University Press, UK, 2003).
\bibitem{ref24}
V. Faraoni, Cosmology in Scalar-Tensor Gravity (Kluwer Academic Publishers, The Netherlands, 2004).
\bibitem{ref25}
B.Bertotti, L. Iess,and P. Tortora, Nature \textbf{425}, 374 (2003).
\bibitem{ref26}
C. M. Will, Living Rev. Rel. \textbf{9}, 3 (2006).
\bibitem{ref27} 
L. Perivolaropoulos, Phys. Rev. D \textbf{81}, 047501 (2010).
\bibitem{ref28}
M. Tegmark et al. (SDSS Collaboration), Phys. Rev. D \textbf{74}, 123507 (2006).
\bibitem{ref29}
V. Acquaviva, C. Baccigalupi, S. M. Leach, A. R. Liddle, and F. Perrotta, Phys. Rev. D \textbf{71}, 104025 (2005).
\bibitem{ref30}
Y.-C. Li, F.-Q. Wu, and X. Chen, Phys. Rev. D \textbf{88}, 084053 (2013).
\bibitem{ref31}
O. Hrycyna, M. Szydlowski and M. Kamionka, Phys. Rev. D \textbf{90}, 124040 (2014).
\bibitem{ref32} 
A. Avilez and C. Skordis, Phys. Rev. Lett. \textbf{113}, 011101 (2014).
\bibitem{ref33}
S. Sen and A. A. Sen, Phys. Rev. D \textbf{63}, 124006 (2001). 
\bibitem{ref34}
M. Arik, M. Calik, M. B. Sheftel, Int. J. Mod. Phys. D \textbf{17}, 225 (2008)
\bibitem{ref35}
R. Garc\'{\i}a-Salcedo, T. Gonz\'{a}lez, and I. Quiros, Phys. Rev. D \textbf{92}, 124056 (2015).
\bibitem{ref36}
S. Weinberg, Gravitation and Cosmology (Wiley, New York, 1972).
\bibitem{ref37}
H. W. Lee, K. Y. Kim, Y. S. Myung, Eur. Phys. J. C \textbf{71}, 1585 (2011)
\bibitem{ref38}
C. Romero, A. Barros, Astrophys. Space Sci. \textbf{192}, 263 (1992).
\bibitem{ref39}
K. Uehara and C. W. Kim, Phys. Rev. D \textbf{26}, 2575 (1982).
\bibitem{ref40}
J. M. Server\'{o} and P. G. Est\'{e}vez, Gen. Rel. Gray. \textbf{15}, 351 (1983). 
\bibitem{ref41}
D. Lorenz-Petzold, Phys. Rev. D \textbf{29}, 2399 (1984).
\bibitem{ref42}
L.O. Pimentel, Astrophys. Space Sci. \textbf{112}, 175 (1984).
\bibitem{ref43}
T. Etoh, M. Hashimoto, K. Arai, S. Fujimoto, Astron. Astrophys. \textbf{325}, 893 (1997).
\bibitem{ref44}
L. E. Gurevich, A. M. Finkelstein, V. A. Ruban, Astrophys. Space Sci. \textbf{22}, 231 (1973).
\bibitem{ref45}
D. A. Tretyakova, A. A. Shatskiy, I. D.Novikov and S. O.Alexeyev, Phys. Rev. D \textbf{85}, 124059 (2012).
\bibitem{ref46} 
T. Singh, T. Singh, J. Math Phys. \textbf{25}, 9 (1984).
\bibitem{ref47} 
A. K. Azad, J. N. Islam, Pramana \textbf{60}, 2127 (2003).
\bibitem{ref48} 
Li. Qiang, G. M. Yong, H. Muxin, Y. Dan, Phys. Rev D \textbf{71}, 061501 (2005). 
\bibitem{ref49} 
M. A. Smolyakov, arXiv: 0711.3811 [gr-qc] (2007).
\bibitem{ref50} 
S. Das and N. Banerjee, Phys. Rev. D. \textbf{78}, 043512 (2008).
\bibitem{ref51}
M. R. Setare \& M. Jamil, Physics Letters B \textbf{690}, 1 (2010)
\bibitem{ref52} 
A. K. Yadav. Research in Astron. Astrophys. \textbf{13}, 772 (2013).
\bibitem{ref53} 
A. T. Ali, A. K. Yadav S.R. Mahmoud, Astrophys Space Sci \textbf{349}, 539 (2014).
\bibitem{ref54}
E. Gaztanaga, E. Garcia-berro , J. Isern, E. Bravo, and I. Dominguez, Phys. Rev. D \textbf{65}, 023506 (2002).
\bibitem{ref55}
J. Ryan, S. Doshi, B. Ratra, MNRAS {\bf 480}, 759 (2018).
\bibitem {ref56}
Moresco, M., et al.: JCAP {\bf05}, 014 (2016)
\bibitem {ref57}
C. Zhang, H. Zhang, S. Yuan, S. Liu, T.-J. Zhang, Y.-C. Sun, Research in Astronomy and Astrophysics, \textbf{14}, 1221 (2014).
\bibitem {ref58}
J. Simon J, L. Verde, R. Jimenez, Phys. Rev. D \textbf{71}, 123001 (2005).
\bibitem {ref59}
M. Moresco M, et al, J. Cosmology Astropart. Phys \textbf{8}, 006 (2012).
\bibitem {ref60}
M. Moresco, et al,J. Cosmology Astropart. Phys \textbf{5}, 014 (2016).
\bibitem {ref61}
A. L. Ratsimbazafy, et al, MNRAS, \textbf{467}, 3239 (2017).
\bibitem {ref62}
D. Stern D, R. Jimenez, L. Verde, M. Kamionkowski, S. A. Stanford, J. Cosmology Astropart. Phys \textbf{2}, 008 (2010).
\bibitem {ref63}
M. Moresco, MNRAS, \textbf{450}, L16 (2015).
\bibitem {ref64}
E. J. Copeland, M. Sami and S. Tsujikawa, Int. J. Mod. Phys. D \textbf{15}, 1753 (2006).
\bibitem {ref65}
W. Hu and N. Sugiyama, Astrophys. J. \textbf{444}, 489 (1995).
\bibitem {ref66}
G. Hinshaw et al. [WMAP Collaboration], Astrophys. J. Suppl. \textbf{208}, 19 (2013).
\bibitem {ref67}
F. Beutler et al., Mon. Not. Roy. Astron. Soc. \textbf{423}, 3430 (2012).
\bibitem {ref68}
A. J. Ross, et al, Mon. Not. Roy. Astron. Soc. \textbf{449}, 835 (2015).
\bibitem {ref69}
S. Alam et al. [BOSS Collaboration], arXiv:1607.03155 [astro-ph.CO].
\bibitem {ref70}
L. Anderson et al. [BOSS Collaboration], Mon. Not. Roy. Astron. Soc. \textbf{441}, 24 (2014).
\bibitem {ref71}
E. A. Kazin et al., Mon. Not. Roy. Astron. Soc. \textbf{441}, 3524 (2014).
\bibitem {ref72}
D. J. Eisenstein et al. [SDSS Collaboration], Astrophys. J. \textbf{633}, 560 (2005).
\bibitem {ref73}
D. J. Eisenstein and W. Hu, Astrophys. J. \textbf{496}, 605 (1998).
\bibitem {ref74}
F. Q. Wu and X. Chen, Phys. Rev. D \textbf{82}, 083003 (2010).
\bibitem {ref75}
J. Muller and L. Biskupek, Classical Quantum Gravity \textbf{24}, 4533 (2007).
\bibitem {ref76}
C. J. Copi, A. N. Davis, and L. M. Krauss, Phys. Rev. Lett. \textbf{92}, 171301 (2004).
\bibitem {ref77}
C. Bambi, M. Giannotti, and F. L. Villante, Phys. Rev. D \textbf{71}, 123524 (2005).
\bibitem {ref78}
D. B. Guenther, L. M. Krauss, and P. Demarque, Astrophys. J. \textbf{498}, 871 (1998).
\bibitem {ref79}
S. E. Thorsett, Phys. Rev. Lett. \textbf{77}, 1432 (1996).
\bibitem {ref80}
R. W. Hellings, et al, Phys. Rev. Lett. \textbf{51}, 1609 (1983).
\bibitem {ref81}
V. M. Kaspi, J. H. Taylor, and M. F. Ryba, Astrophys. J. \textbf{428}, 713 (1994).
\bibitem {ref82}
K.-C. Chang and M. C. Chu, Phys. Rev. D \textbf{75}, 083521 (2007).
\bibitem {ref83}
G. Chen, B. Ratra, B, PASP \textbf{123}, 1127 (2011).
\bibitem {ref84}
E. Aubourg, et al, Phys. Rev D \textbf{92}, 123516 (2015).
\bibitem {ref85}
G. Chen, S. Kumar, B. Ratra, B, Astrophys. J. \textbf{835}, 86 (2017).
\bibitem {ref86}
G. Hinshaw, et al, Astrophys.J.Suppl.Ser \textbf{208}, 25 (2013).
\bibitem {ref87}
H. Amirhashchi, Phys. Rev D \textbf{97}, 063515 (2018).
\bibitem {ref88}
Nisha Rani et al, JCAP. \textbf{12}, 045 (2015).
\bibitem {ref89}
M. Vargas dos Santosa, R. R. R. Reisa and I. Wagaa, JCAP. \textbf{02}, 066 (2016).
\bibitem {ref90}
H. Amirhashchi, arXiv: 1811.05400 [astro-ph.CO] (2018).
\bibitem {ref91}
H. Yu, B. Ratra and Fa-Yin, Wang, Astrophys. J. \textbf{856}, 3 (2018).
\bibitem {ref92}
R. Goistri, et al, JCAP \textbf{03}, 027 (2012).
\bibitem {ref93}
S. Capozziello, O. Luongo, E.N. Saridakis, Phys. Rev. D \textbf{91}, 124037 (2015).
\bibitem {ref94}
H. Amirhashchi, S. Amirhashchi, Phys. Rev. D \textbf{99}, 023516 (2019).
\end{thebibliography}
\end{document}